\documentclass[twocolumn,aps,prl,showpacs]{revtex4}
\usepackage[dvips]{graphicx}
\usepackage{amsmath}
\usepackage{amsfonts}
\usepackage{color}

\begin{document}

\title{The multiphysics of prion-like diseases: progression and atrophy}

\author{Johannes Weickenmeier$^\dagger$,  Ellen Kuhl$^{\ddagger}$, and Alain Goriely$^*$}

\affiliation{\it $^\dagger$Stevens Institute of technology, New Jersey, USA\\ $^\ddagger$Living Matter Laboratory, Stanford University, Stanford USA,\\
$^*$Mathematical Institute, University of Oxford, Oxford UK .}

\begin{abstract} 
Many neurodegenerative diseases are related to the propagation and accumulation of toxic proteins throughout the brain.  The lesions created by aggregates of these toxic proteins further lead to cell death and accelerated tissue atrophy.  A striking feature of some of these diseases is their characteristic pattern and evolution, leading to well-codified disease stages visible to neuropathology and associated with various cognitive deficits and pathologies. Here, we simulate  the anisotropic propagation and accumulation of toxic proteins in full brain geometry. We show that the same model with different initial seeding zones reproduces the characteristic evolution of different prion-like diseases. We also recover the expected evolution of the total toxic protein load. Finally, we couple our transport model to a mechanical atrophy model to obtain the typical degeneration patterns  found in  neurodegenerative diseases.
 \end{abstract}

\maketitle

\noindent\textbf{Introduction:} Age-related neurodegenerative disorders are extremely complex and multifaceted pathologies. Yet, their evolution is known to be closely associated with the progression of particular protein aggregates. When these proteins misfold and/or aggregate they can be transported into the brain tissue and become the seed for further misfolding and aggregation \cite{jucker2013self}. Unless these toxic proteins are removed or stabilized,  this chain-reaction proceeds. As toxic proteins invade the brain and form larger aggregates, they prevent the proper function of other proteins and create tissue lesions. over time, it leads to a disruption of the proper function of the nervous system, a loss of tissue structure, necrosis, brain atrophy, and ultimately death \cite{brettschneider2015spreading}. 

In prion diseases, such as  Creutzfeldt-Jakob disease, the infectious agents are typically small soluble misfolded isoforms of proteins that can aggregate and  trigger the misfolding of the same protein found in its benign form.  When these aggregates are large enough, they form characteristic tissue lesions.
A similar mechanism of protein corruption and propagation is  found in both secondary injury following traumatic brain injury  \cite{mckee2013spectrum,daneshvar2015post,cruz2017pathological} and  common age-related neurodegenerative diseases such as Alzheimer's disease, the most common form of dementia \cite{watts2014serial}. For instance, it is known from \textit{in vitro} experiments that the amyloid-$\beta$ protein can form aggregates that propagate through the tissues. These aggregates have a distribution of sizes from small soluble groups to large insoluble fibrils. Their accumulation as plaques is believed to play a key role  in Alzheimer's disease.  Tau proteins also play an important role in these diseases \cite{goedert2015alzheimer}. These small proteins are known to stabilize microtubules within the axon. Hyperphosphorylated tau proteins can act as seeds for  further misfolding and aggregation. These aggregates are found within the axon where they are rapidly transported through the usual interneuronal transport mechanisms \cite{bressloff2013stochastic}. They are also transported from the cell to the extracellular space where they diffuse through secretion and  damage of the host cell, and from the extracellular space into cells by endocytosis \cite{soto2012vivo,wu2013small}. These aggregates are subject to biological clearance slowing  down or arresting  the overall spreading process. The particular role of neuronal pathways in this transport mechanism may provide an explanation for the observation that the spreading of neurofibrillary tangles 
follows a characteristic topographic pattern.

 \begin{figure}[h]
\centering
\includegraphics[width=0.99\columnwidth]{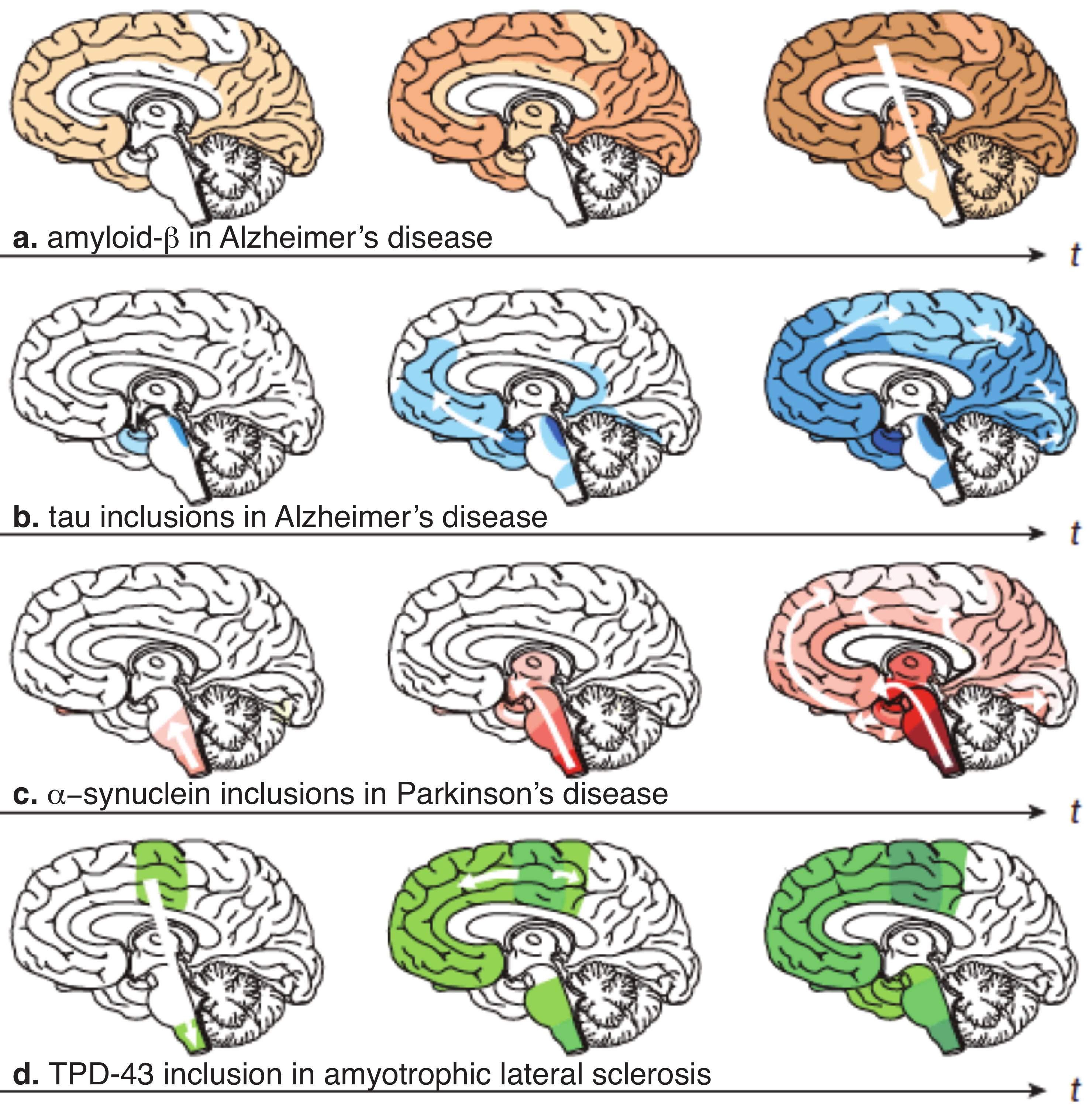}
\caption{Typical spatial progression of protein aggregates in various neurodegenerative diseases. From Jucker \cite{jucker2013self}.}\label{Fig1-Jucker}
\end{figure}
 Independently of their molecular origin or specific actions, these pathologies, known as \textit{prion-like diseases}, share the same macroscopic features when viewed as a spatio-temporal evolution process \cite{walker2015neurodegenerative}. The progression is believed to include the following steps: \textit{(i)} seeding of misfolded proteins, \textit{(ii)} templated misfolding and aggregation of native (homologous) proteins, \textit{(iii)} growth of aggregates, \textit{(iv)} fragmentation and spatial propagation of aggregates of different sizes, \textit{(v)} assembly of misfolded proteins into secondary structures (e.g. protofibrils and fibrils), and \textit{ (vi)} formation of larger tertiary structures (lesions) at the tissue level. A striking feature of each particular pathology is its systematic evolution pattern. We posit that the main reason for such reproducibility is that the overall progression of the disease is governed by basic transport processes and that the key difference between individual diseases, aside from the particular set of proteins involved, is the original location of the seeds, leading to  a characteristic spatio-temporal evolution (Fig. 1) with respective pathological symptoms. 

An essential feature of the dynamics  is that small proteins are preferentially transported transynaptically but can also diffuse from the extracellular space to the cells through endocytosis \cite{frost2009propagation}. These effects lead to fast transport along neuronal pathways and a slower diffusion in directions perpendicular to the axon bundles. Here, we couple an anisotropic transport process with a mechanical model to take into account the relative effect of seed location, anisotropy, and convoluted brain geometry on the damage patterns and the resulting atrophy of brain tissue \cite{gogeho15,gobuku15}. We show that a minimal model for transport and atrophy is sufficient to recover the main qualitative spatial features of different neurodegenerative diseases as well as the overall increase of key biomarkers.\\
%

\noindent\textbf{The model:} 
We couple a transport model to a mechanical model of the brain. The geometry of the brain is taken to be a connected domain $\Omega$ in either $\mathbb{R}^{2}$ (slice simulation) or $\mathbb{R}^{3}$ (full brain simulation) composed of gray ($\Omega_\text{gray}$) and white ($\Omega_\text{white}$) matters with different material properties and total volume (or area) $V$. We first consider the transport process. Different models for the aggregation/propagation process have been proposed ranging from graph Laplacian diffusion on simplified structural networks \cite{abdelnour2014network} to Smoluchowski-type equations \cite{bressloff2014waves,bertsch2016alzheimer} tracking the aggregate-size distribution. Here, we use a minimal model for the propagation of a toxic protein that captures the important physical characteristics of the problem: transformation of existing native proteins and anisotropic diffusion along axonal pathways. We assume that the concentration $c(\mathbf{x},t)\in\Omega_\text{white}$  follows  the Fisher-KPP equation
\begin{equation}
\frac{\partial c}{\partial t}=\nabla\cdot(\textbf{D}\nabla c)+\alpha c (1-c),\label{Fischer}
\end{equation}
where $\textbf{D}=d_{\perp} \mathbf{1}+(d_{\parallel}-d_{\perp}) \boldsymbol{\gamma}\otimes \boldsymbol{\gamma}$ 
is a transversely anisotropic diffusion tensor chosen to have a preferential direction along the axon bundle characterized by the unit vector $\boldsymbol{\gamma}=\boldsymbol{\gamma}(\mathbf{x},t)$. Here, $\alpha>0$,   $d_{\perp}$ is the regular tissue diffusion and $d_{\parallel}\gg d_{\perp}$ is the  diffusion along the axons.  This model  describes the overall increase of toxic protein concentration, assuming that the pool of protein in the native form is sufficiently large as not to be affected by the increase in toxic protein. The nonlinearity provides a saturation term expressing the maximal concentration of toxic proteins achievable (taken to be 1 without loss of generality). In one dimension (along the axonal bundles), the system supports traveling fronts with a velocity $v=2 \sqrt{\alpha d_{\parallel} }$ leading to a complete invasion in a characteristic time $T=L/v$, where $L$ is the typical length scale for the entire brain, which is on the order of 16cm for human brains. In the gray matter, for $\mathbf{x}\in_\text{gray}$, we assume simple isotropic diffusion from cells to cells different from the white matter,  (Eq. (1) with $d_{\parallel}=d_{\perp}$ and $\alpha=0$).

We are  interested in tracking two  quantities related to the concentration field $c(\mathbf{x},t)$. First, the  time $\tau(\mathbf{x},C_{\text{crit}})$ at which the concentration first reaches a critical level close to the saturation level $C_{\text{crit}}=(1-\varepsilon)$ 
defined implicitly by 
$c(\mathbf{x},\tau)=C_{\text{crit}}$. Second, we investigate the temporal evolution of the average amount of toxic protein as a possible biomarker, computed as 
\begin{equation}
C(t)=\frac{1}{V} \int_{\Omega} c(\mathbf{x},t)\, \text{d}\mathbf{x}.
\end{equation}
 
We couple the transport process to the mechanics of atrophy by assuming that gray and white matter tissues are morphoelastic materials \cite{ku13,goriely17} and assigning an isotropic shrinking factor $0<\vartheta<1$ at each point depending on the concentration of toxic proteins. Given an initial  reference configuration, atrophy is characterized by a deformation $\boldsymbol{\varphi}:\mathcal{B}_{0} \to \mathcal{B}_{t}$,\ $\textbf{x}(\textbf{X},t)=\boldsymbol{\varphi}(\textbf{X},t)$   from the reference brain $\mathcal{B}_{0}$  to the aged brain after shrinking $\mathcal{B}_{t}$. Here, $\textbf{x}(\textbf{X},t)$ is the position at time $t$ of the material point originally  located at $\textbf{X}$ at time $t=0$. This deformation is obtained by first defining the deformation gradient $\mathbf{F}=\nabla_{\mathbf{X}}\boldsymbol{\varphi}$ and assuming that it can be decomposed as $\mathbf{F}=\vartheta^{1/3} \mathbf{A}$, where $\vartheta$ denotes the volumetric change ($\vartheta<1$ represents shrinking and $\vartheta>1$ would correspond to growth).  The elastic deformation tensor  $\mathbf{A}$, is computed by solving the Cauchy equation div$(\textbf{T})=\textbf{0}$ for the Cauchy stress tensor $\mathbf{T}=J^{-1} \mathbf{A}\cdot \partial_{\mathbf{A} } W$  associated with the strain-energy density $W=W(\mathbf{A},\textbf{x})$ and $J=\text{det}(\mathbf{A})$.  The amount of shrinking is proportional to the total exposure of the tissue to the toxic proteins, so that  $\partial_t\vartheta=\delta c$, where $\delta>0$ is an overall parameter that describes the rate of removal of material. \\


\noindent\textbf{Computational investigation:}
   We create three finite element models from T2-weighted magnetic resonance images of a 32-year old male using Simpleware: a two-dimensional sagittal model, a two-dimensional coronal model, and a fully three-dimensional, anatomically accurate whole brain model. The sagittal model consists of 13442 linear triangular elements and 7216 nodes; the coronal model consists of 38223 linear triangular elements and 19813 nodes; and the three-dimensional whole brain model consists of 401940 linear tetrahedral elements and 80233 nodes. To model anisotropic toxic protein propagation, we register the axonal fiber orientation from the diffusion tensor magnetic resonance image of the subject and create an element based fiber-orientation map. 
   
For the initial seeding, we assume a concentration $c(\mathbf{x})=1$ for all $\mathbf{x}\in\Omega_{\text {seed}}\subset\Omega$ and $c(\mathbf{x})=0$ otherwise.
Then, using Eq.~(\ref{Fischer}) we propagate the toxic protein in space and time across the brain with a diffusivity contrast of $d_{\parallel} : d_{\perp} = 100$ to ensure a faster propagation along the axonal fiber direction. In the white matter, we use $d_{\parallel}=100$\,mm$^{2}$/year and $\alpha=0.5/$year. In the  gray matter, we use   $d_{\parallel}=d_{\perp} =10$\,mm$^{2}$/year.
  
  We post-process the concentration field $c$ to extract the activation time $\tau(\mathbf{x},C_{\text{crit}})$. From those activation times, we create activation maps of all four cases of prion-like disease similar to Fig.~\ref{Fig1-Jucker}. Using Eq. (2), from the spatio-temporal concentration field $c(\mathbf{x},t)$, we extract the integration of the toxic protein concentration in time $C(t)$ to create a temporal map of the toxic protein concentration averaged over the entire brain \cite{jack2013biomarker}. 
 \begin{figure}[ht]
\centering
\includegraphics[width=0.99\columnwidth]{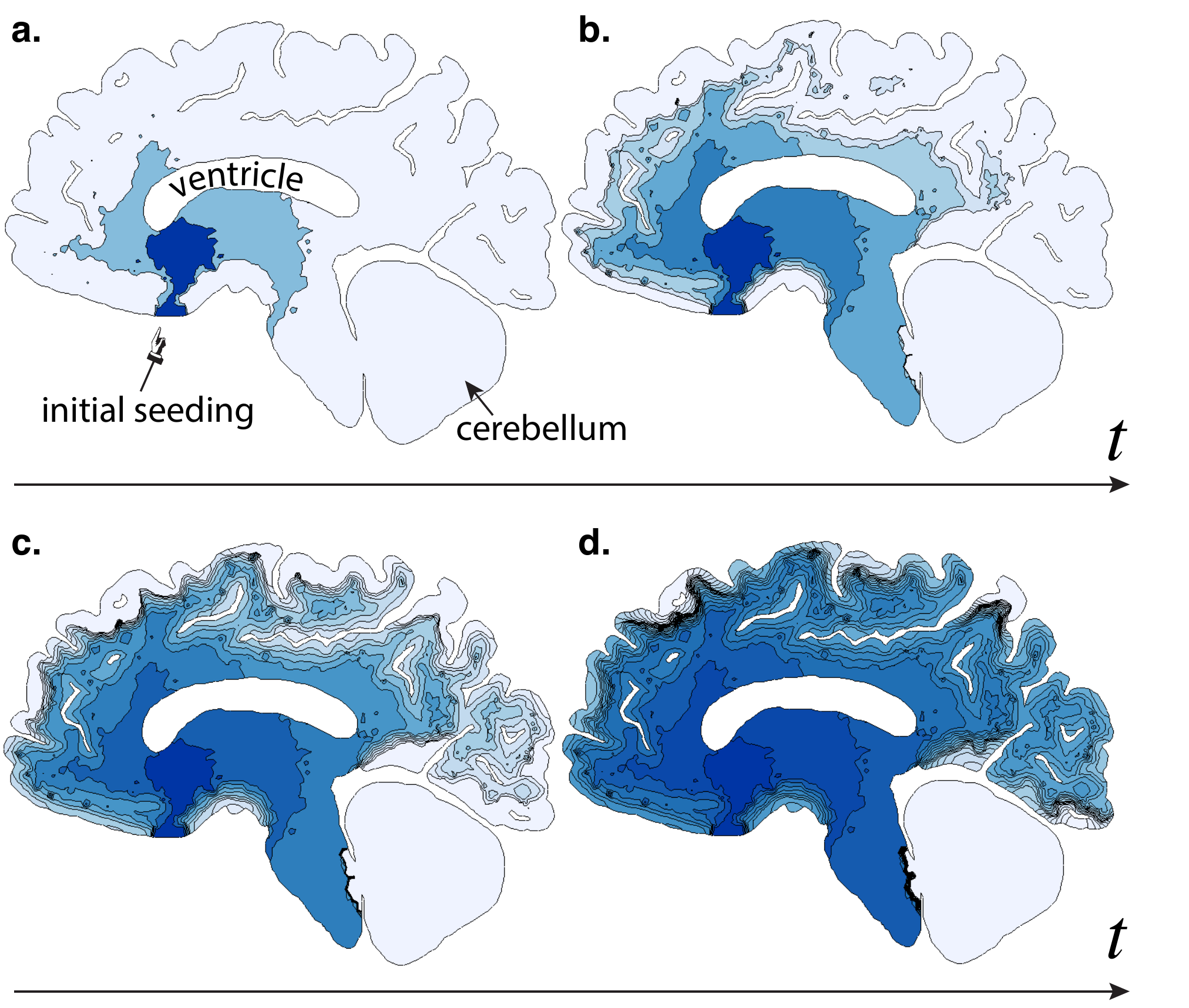}
\caption{Activation maps of tau inclusions in brains of patients with Alzheimer's disease in a 2D sagittal slice.
}\label{Fig-Jucker2D}
\end{figure} 
  
  Finally, using the tissue atrophy model with different atrophy rates $\delta$ in the gray and white matter tissue, we utilize our two-dimensional coronal model to create atrophy maps. It is important to note that atrophy takes  place primarily in locations with elevated concentrations of toxic proteins. Therefore,  the change of volume does not significantly affect the time evolution of the concentration since it is close to maximal at the locations where atrophy takes place.  We take advantage of this property to compute tissue shrinking as a post-processing step based on the values of the concentration at different time points.  Since we assume that after each step of shrinking, there is total relaxation of the stresses,  the mechanical stresses generated are small during shrinking and we can use a standard compressible neo-Hookean model  valid in small deformations with parameters fitted to experimental data: $W=c_{1}(\text{tr}(\mathbf{A}^{\text{T}}\cdot\mathbf{A})-3)+c_{2}(J-1)^2$ with $c_{1}=666.56\,\text{Pa}$ and $c_{2}=1777.5$ Pa 
  \cite{mihai2015comparison}. For the relative shrinking rate, it is known that the gray matter tissue shrinks looses more volume than the white matter tissues \cite{thompson2003dynamics}. For our computation, we use a ratio $\vartheta_\text{gray}: \vartheta_\text{white}=4:1$.\\
%


%

\noindent\textbf{2D simulations:}  We start with a two-dimensional simulation on a sagittal slice (Fig.~\ref{Fig-Jucker2D}). 
 The region  includes the ventricle and cerebellum but, through no-flux boundary conditions,  it is assumed that no progression takes place into these regions. Based on the description of the evolution of tau inclusions in Alzheimer's disease, we simulate the propagation of toxic tau proteins by considering an initial seeding close to the entorhinal cortex (Fig.~\ref{Fig-Jucker2D}a).  We observe the quick  progression into the brain stem and  the hippocampus (Fig.~\ref{Fig-Jucker2D}a), to paralimbic and adjacent medial-basal temporal cortex (Fig.~\ref{Fig-Jucker2D}b), to cortical association areas (Fig.~\ref{Fig-Jucker2D}c), and eventually reaching primary sensory-motor and visual areas. The relatively rapid progression along the hippocampus is associated with a strong anisotropy along the axonal pathways tangent to the ventricles. This progression is consistent with the well-established  Braak stages of Alzheimer's disease \cite{braak1991neuropathological}: Stage I-II \textit{transentorhinal} (Fig.~\ref{Fig-Jucker2D}a); Stage III-IV \textit{limbic} (Fig.~\ref{Fig-Jucker2D}b); Stage V-VI \textit{isocortical}  (Fig.~\ref{Fig-Jucker2D}c-d).\\
\begin{figure}[h]
\centering
\includegraphics[width=0.99\columnwidth]{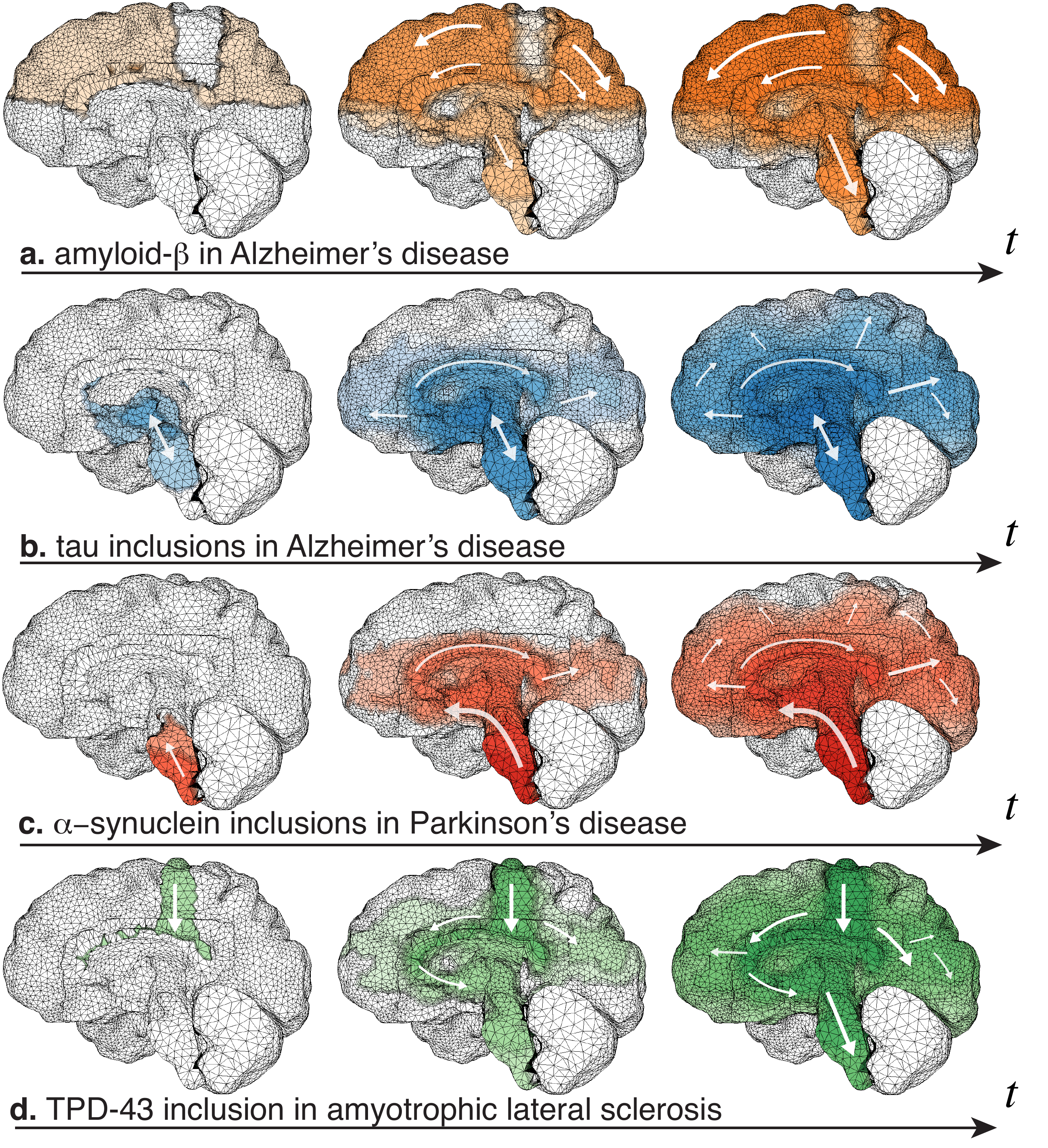}
\caption{Simulated activation maps for the spatial progression of toxic protein in full brain geometry  for various initial seeding regions matching Fig. 1~\cite{jucker2013self}. 
}\label{Fig-Jucker3D}
\end{figure}  \\

\noindent\textbf{3D simulations:} The previous analysis uses a high-resolution mesh but lacks the  three-dimensional geometry of the brain. Next, we use the accepted knowledge about different neurodegenerative diseases to compare the progression of various toxic proteins. To test the specific role of geometry in the progression, we assign the same parameter values for all cases and only modify the initial seeding region of the four cases discussed in \cite{jucker2013self} and depicted in Fig.~\ref{Fig1-Jucker}. For every individual seeding region we show, in Fig.~\ref{Fig-Jucker3D},  the time $\tau$ to reach a critical level of damage.

We observe generic  progression trends in all cases: \textit{(i)} as soon as a sufficient level of toxic proteins reaches the area close to the ventricles, there is a fast progression in the limbic system (around the ventricles) leading to a rapid invasion of the temporal and occipital lobes; \textit{(ii)} once a sufficient level of toxic protein is reached within the cortex, further invasion through the cortex takes place; \textit{(iii)} if the parietal lobe is not directly involved initially, it only becomes invaded in the last stages of the disease. 

\begin{figure}[h]
\centering
\includegraphics[width=1\columnwidth]{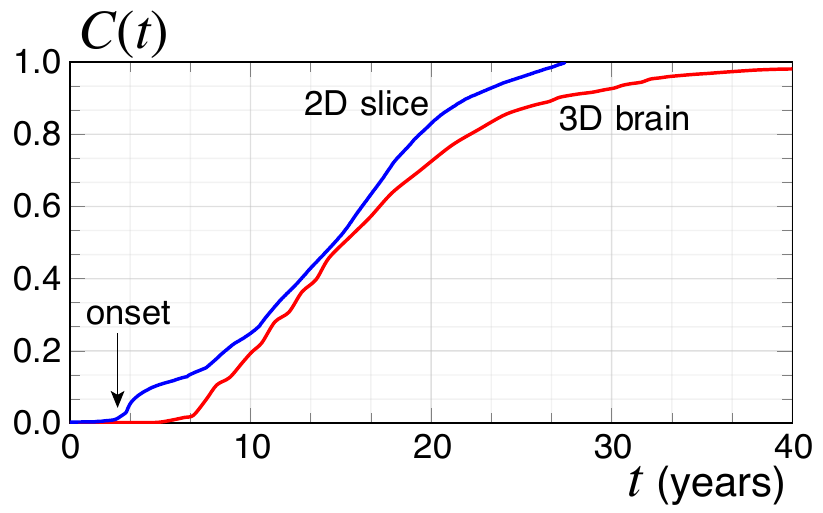}
\caption{Evolution of the total concentration for the simulation of Fig 2 (2D sagittal slice) and of Fig 3b (3D tau concentration in Alzheimer's disease). 
}\label{Fig4-Jack}
\end{figure}
An indirect way to follow the evolution of neurodegenerative diseases is to look at the averaged concentration of toxic proteins or  associated biomarkers. The paradigm for this evolution is based on the so-called \textit{Jack's curves}. These curves are the expected, mostly hypothetical, evolution of a typical biomarker concentration as a function of time \cite{jack2013tracking}. They take the shape of sigmoid-like functions: a slow evolution at first during incubation, followed by a sharp increase during the outbreak and leading to saturation when the disease is fully established \cite{jack2013biomarker}. The variation of the typical parameters entering the sigmoids are used to represent physiological quantities such as  time of outbreak, and rate of spreading. These curves  serve as a general guideline to understand the variations between individuals, pathologies, and the effect of various stimuli or therapeutics  \cite{van2017molecular}. We can  extract these curves from  the simulation using (2) as shown in Fig.~\ref{Fig4-Jack}. 
Our simulations recover the predicted behavior for this type of disease integrated across the entire brain.   This profile is not unexpected. Indeed, we known that Equation~(\ref{Fischer}) supports fronts in one-dimension that have  profile and similar  evolution equations for Alzheimer's diseases. A simple estimate using the typical wave speed of front propagation in one-dimension leads to a lower and upper bound for the total invasion of 11.3 and 35.8 years, respectively, obtained by using the front velocity with either the fast axonal diffusion constant $d_{\parallel}\approx 100\,\text{mm}^{2}/\text{year}$ or the slow gray matter diffusion constant $d\approx 10\,\text{mm}^{2}/\text{year}$.  These estimates are consistent with the characteristic time scales observed in Fig.~\ref{Fig4-Jack}.  Similar curves have also been obrained on 2D slices in different geometries~\cite{bertsch2016alzheimer}. Here, we observe that the fully 3D evolution on an actual brain model also follows the same general trend of a progressive invasion. \\

\begin{figure}
\centering
\includegraphics[width=0.99\columnwidth]{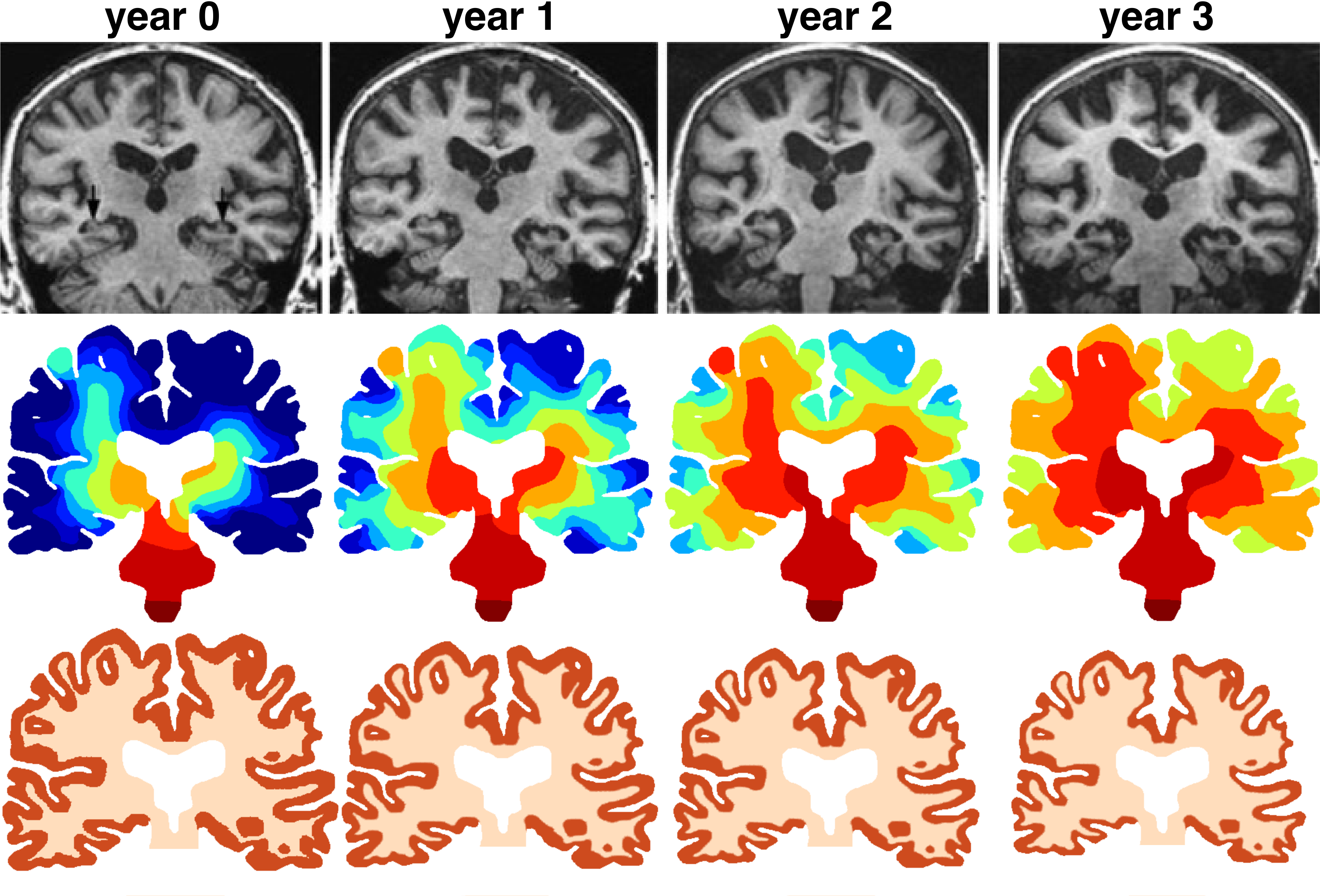}
\caption{Atrophy. Top row: magnetic resonance images showing hippocampal atrophy in AlzheimerÕs disease \cite{lehericy2007magnetic}: yearly examination of the same subject is shown. Increasing hippocampal and atrophy is observed together with ventricular enlargement and widening of cortical sulci. Middle row: activation time based on initial seeding in the brain stem. Bottom row: shrinking based on the activation time and toxic protein concentration with a ratio $\vartheta_\text{gray}: \vartheta_\text{white}=4:1$.}\label{Fig5-shrink}
\end{figure}

\noindent\textbf{Shrinking:}  We now turn to the effect of elevated toxic proteins on the tissue. It is known that the formation of large aggregates or high concentration of toxic proteins prevents the proper function of neuronal cells, leading to ischemia, and eventually tissue removal. For instance, in the top row of Fig.~\ref{Fig5-shrink}, there is a marked and rapid atrophy associated with the disease. We use our propagation model on a similar coronal slice to compute the activation time at three different time points (middle row). For each of these time points we compute the shrinking of the slice and show the change in geometry from the previous one (bottom row).  The degeneration-induced atrophy patterns (bottom row) agree well with the atrophy pattern observed in Alzheimer's disease (top row).  
\\

\noindent\textbf{Conclusions:}  Neurodegenerative diseases are known for their extreme complexity and medical science has wrestled with this major challenge by amassing a considerable amount of information despite few successes. One of the remarkable features that appear in these diseases is their remarkable reproducible topographic propagation pattern. Each disease has a well-defined spatio-temporal evolution that has been documented over the years and catalogued into stages, each associated with typical symptoms. This evolution appears much more controlled than other diseases such as cancer and has led to three main hypotheses in recent years. First, it has been proposed that  diseases share the same characteristics as prion diseases and are based on the seeding, propagation, and accumulation of toxic proteins involved in regular sub-cellular functions such as alpha-synucleins or tau proteins. Second, it has been proposed that the overall evolution of a typical biomarker follows regular sigmoid-like curves. Third, a given disease is  associated with typical atrophy patterns.

There have been multiple mathematical models proposed for neurodegenerative diseases, but most of them focus on biochemical pathways, cellular interactions, and  the formation of amyloids \cite{zamparo2010simplified,michaels2016hamiltonian,lloret2017impact}. Our approach here is radically different as we study the problem at the largest possible scale of the brain.  Our minimal model takes into account the main microscopic features of these diseases, but incorporated the full brain geometry and axonal directions from diffusion tensor imaging. Despite the complexity of these diseases our model recovers the three aforementioned key features of neurodegenerative disease progression and suggests that  brain geometry, transport anisotropy, and mechanics play a central role in neurodegeneration and atrophy pattern formation. \\



 \begin{acknowledgments}
This work was supported by the NSF Grant CMMI 1727268 to Ellen Kuhl.  The support for  Alain Goriely by the Engineering and Physical Sciences Research Council of Great Britain under research grant EP/R020205/1 is gratefully acknowledged.
\end{acknowledgments}


\end{document}